\documentclass[twoside,11pt]{article}
\usepackage{a4wide,cite,latexsym,amsfonts,amssymb,exscale,epsfig,bbm}
\usepackage[centertags,sumlimits,intlimits,namelimits,reqno]{amsmath}

\usepackage{amsthm}
\theoremstyle{definition}

\newtheorem{theorem}{Theorem}[section]
\newtheorem{lemma}[theorem]{Lemma}
\newtheorem{corollary}[theorem]{Corollary}
\newtheorem{proposition}[theorem]{Proposition}
\newtheorem{remark}[theorem]{Remark}
\newtheorem{definition}[theorem]{Definition}

\def\draft#1{}%
\def\conts{}%
 
\renewcommand{\theequation}{\thesection.\arabic{equation}}%
\makeatletter
\@addtoreset{equation}{section}
\makeatother

\newcounter{mathletter}%
\newcommand{\bmathletter}{%
  \refstepcounter{equation}%
  \setcounter{mathletter}{\value{equation}}%
  \setcounter{equation}{0}%
  \renewcommand{\theequation}{%
    \mbox{\thesection.\arabic{mathletter}\alph{equation}}}}%
\newcommand{\emathletter}{\setcounter{equation}{\value{mathletter}}}%
\newenvironment{mathletters}{\bmathletter}{\emathletter}%

\newenvironment{myenumerate}{%
  \begin{enumerate}
  \setlength{\itemsep}{0pt}
  \setlength{\parskip}{0pt}}{\end{enumerate}}

\def\openone{\mathbbm{1}}%
\def\keywords#1{\noindent key words: #1\par}%
\def\acknowledgements{\section*{Acknowledgements}}%

\def\dopreprint{\hfill{\small\thepreprint}\\}%
\def\preprint#1{\def\thepreprint{#1}}%

\def\ket#1{\left|#1\right>}

\def\sym#1{{\mathcal #1}}
\def\emph#1{{\sl #1\/}}

\let\phi=\varphi
\let\theta=\vartheta
\let\epsilon=\varepsilon

\let\hat=\widehat
\let\tilde=\widetilde

\def\diag{\mathop{\rm diag}\nolimits}

\def\id{\mathop{\rm id}\nolimits}

\def\id{\mathop{\rm id}\nolimits}
\def\Span{\mathop{\rm span}\nolimits}
\def\max{\mathop{\rm max}\nolimits}
\def\min{\mathop{\rm min}\nolimits}
\def\qdet{\mathop{\rm qdet}\nolimits}

\def\address#1{\date{{\sl #1}\\\ \\\theversion}\gdef\date##1{}}%
\def\version#1{\gdef\theversion{#1}}%

\makeatletter
\newfont{\@aidxte}{cmsy10}
\newfont{\@aidxel}{cmsy10 scaled 1095}
\newfont{\@aidxtw}{cmsy10 scaled 1200}
\newlength\@aidxtexvi
\newlength\@aidxtexvii
\newlength\@aidxelxvi
\newlength\@aidxelxvii
\newlength\@aidxtwxvi
\newlength\@aidxtwxvii
\newcommand{\alignidx}[1]{%
  \@aidxtexvi=\fontdimen16\@aidxte
  \@aidxtexvii=\fontdimen17\@aidxte
  \@aidxelxvi=\fontdimen16\@aidxel
  \@aidxelxvii=\fontdimen17\@aidxel
  \@aidxtwxvi=\fontdimen16\@aidxtw
  \@aidxtwxvii=\fontdimen17\@aidxtw
    {\mbox{$%
    \fontdimen16\@aidxte=2.9pt
    \fontdimen17\@aidxte=2.9pt
    \fontdimen16\@aidxel=3.1pt
    \fontdimen17\@aidxel=3.1pt
    \fontdimen16\@aidxtw=3.3pt
    \fontdimen17\@aidxtw=3.3pt
    #1$}}%
    \fontdimen16\@aidxte=\@aidxtexvi
    \fontdimen17\@aidxte=\@aidxtexvii
    \fontdimen16\@aidxel=\@aidxelxvi
    \fontdimen17\@aidxel=\@aidxelxvii
    \fontdimen16\@aidxtw=\@aidxtwxvi
    \fontdimen17\@aidxtw=\@aidxtwxvii}
\makeatother

\def\eqref#1{(\ref{#1})}%

\def\g{{\mathfrak{g}}}
\def\gl{{\mathfrak{gl}}}
\def\ssl{{\mathfrak{sl}}}

\def\ie{{\sl i.e.\/}}
\def\eg{{\sl e.g.\/}}

\def\etc{{\sl etc.\/}}

\def\nn{\notag}

\def\C{{\mathbbm C}}
\def\N{{\mathbbm N}}

\version{28 June 2000}%
\preprint{DAMTP-2000-63}%

\def\tp{{\rm tr}}
\def\Ha{H\otimes\openone}
\def\Hb{\openone\otimes H}
\def\Dop{\Delta^{\rm op}}

\begin{document}
%

\title{\dopreprint Factorizing twists and the universal $R$-matrix of
  the\\ Yangian $Y(\ssl_2)$}
\author{Hendryk Pfeiffer\thanks{e-mail: H.Pfeiffer@damtp.cam.ac.uk}}
\address{Department of Applied Mathematics and Theoretical Physics,\\
  Centre for Mathematical Sciences, Cambridge, UK}
\date{\version}
\maketitle

%
\begin{abstract}
%

  We give an explicit construction of the factorizing twists for the
  Yangian $Y(\ssl_2)$ in evaluation representations (not necessarily
  finite-dimensional). The result is a universal expression for the
  factorizing twist that holds in all these representations. The
  method is general enough to recover the universal $R$-matrix of
  $Y(\ssl_2)$ up to its character in the form specialized to generic
  evaluation representations. The method presented here is
  particularly amenable to generalizations because it involves
  only elementary operations applied to representations of the
  Yangian.

\end{abstract}

\conts

%
\section{Introduction}
%

Drinfel'd twists~\cite{Dr90} have been applied very successfully to
quantum integrable spin chains in the framework of the Algebraic Bethe
Ansatz. It was shown by Maillet and Sanchez de Santos in~\cite{MaSa96}
that the Yangian $Y(\ssl_2)$ and the quantized envelope of the affine
Lie algebra $U_q(\hat\ssl_2)$ admit factorizing twists in at least the
fundamental evaluation representation. Such a factorizing twist $F$
equips the quantum algebra with a new coproduct
$\Delta_F=F\cdot\Delta\cdot F^{-1}$ which is cocommutative. In
representations the effect is that tensor products of representations
admit a change of basis such that all expressions originating from
coproducts are symmetric under exchange of the tensor factors. The
cocommutativity of the coproduct has provided a dramatic
simplification of the Algebraic Bethe Ansatz. The application of
factorizing twists has triggered an interesting development with
far-reaching new results for correlation
functions~\cite{KiMa99,IzKi99,KiMa00}.

From a mathematical point of view these twists are particularly
interesting because, if they exist, they factorize the universal
$R$-matrix, $R=\alignidx{F_{21}^{-1}\cdot F_{12}}$ (hence the name),
and thus can be considered more fundamental than the $R$-matrix
itself. We demonstrate in this paper that, for evaluation
representations, the twists can even be used to determine the
universal $R$-matrix up to a factor. On the other hand, the
factorizing twists provide the Yangian $Y(\ssl_2)$ with an additional,
very restrictive structure which has not been fully exploited in the
analysis of the algebra yet.

Recently factorizing twists have been found for the evaluation
representation of the Yangian $Y(\ssl_n)$ corresponding to the first
fundamental representation of $\ssl_n$~\cite{AlBo00} and for all
finite-dimensional evaluation representations of the Yangian
$Y(\ssl_2)$~\cite{Te99}, though the diagonal part of the twist was not
determined there. Here the term `diagonal part' refers to the Gauss
decomposition of the twist $F=F_0F_-$ into a diagonal operator $F_0$
and a lower triangular $F_-$. Likewise the universal $R$-matrix
can be Gauss decomposed $R=\alignidx{R_+\,R_0\,R_-}$.

It has been known for some time~\cite{ChPr94} that the Yangian
$Y(\ssl_2)$ is pseudo triangular. In view of the results for
finite-dimensional semi-simple Hopf algebras~\cite{EtGe98a} it is
natural to conjecture that the pseudo-triangularity of $Y(\ssl_2)$
might lead to the construction of factorizing twists at least in a
certain class of representations. This hypothesis is supported by the
results of~\cite{Te99}, but there it was not shown explicitly how the
twists factorize the (pseudo-)universal $R$-matrix.

The study of the factorizing twists in~\cite{Te99} used a modification
of the Functional Bethe Ansatz.
This method has the advantage that the calculation of the
iterated twisted coproduct is manageable and that this coproduct is
cocommutative by construction. Cocommutativity is actually a property
of the polynomial interpolation which is employed in the Functional
Bethe Ansatz. One of the disadvantages, however, is that the change of
basis from the non-cocommutative situation to the cocommutative one,
\ie\ the factorizing twist in a particular representation, is given
only implicitly. In~\cite{Te99} it was possible to identify the
triangular part of the twist and to show that it agrees with the
triangular part of the Gauss decomposition of the $R$-matrix of
$Y(\ssl_2)$ which is `universal' for all evaluation representations of
$Y(\ssl_2)$~\cite{KhTo96}, but the diagonal part was not given
explicitly.

In this paper, we present a conceptually very simple direct and
completely explicit calculation of the factorizing twist for
evaluation representations of $Y(\ssl_2)$ which are modelled using
highest weight representations of the corresponding Lie algebra
$\ssl_2$ which are not necessarily finite-dimensional. It allows us to
show in detail how the twist factorizes the triangular and diagonal
parts of the universal $R$-matrix on evaluation representations. This
complements the result of~\cite{Te99} and confirms the conjecture that
the factorizing twists come from a universal object. This is the first
main result of this paper.

From our construction of the factorizing twist, it is furthermore
possible to recover the generic form of the $R$-matrix for evaluation
representations up to a scalar factor, the so-called
character~\cite{KhTo96}, which depends on the given
representations. Our method allows us to reproduce this `universal'
expression for the $R$-matrix which was derived in~\cite{KhTo96} where
the quantum double $\sym{D}Y(\ssl_2)$ of the Yangian was used to
calculate it. The $R$-matrix obtained via the factorizing twist
appears furthermore in a more natural form that is simplified compared
with the expression derived via the quantum double.

Obviously, once the $R$-matrix is known for evaluation
representations, the axioms of the quasi-triangular structure
determine the $R$-matrix for tensor products of evaluation
representations. Hence our method is sufficient to obtain $R$-matrices
for all tensor products of evaluation representations as well. These
cover in particular all finite-dimensional irreducible representations
of $Y(\ssl_2)$ and those representations which are of interest in
applications to integrable systems.

We wish to emphasize that the direct method presented here does not
make use of any fusion methods, but rather determines the factorizing
twist and the $R$-matrix in one step for all evaluation
representations. The ideas employed here are simple enough to offer a
chance for a generalization both to the trigonometric case where the
algebra is $U_q(\hat\ssl_2)$ and also to the cases of higher rank.

The paper is organized as follows. In Section~\ref{sect_prep}
we recall the basic properties of factorizing twists and fix our
notation for the study of evaluation representations of the Yangian
$Y(\ssl_2)$.

In Section~\ref{sect_triangular} we begin the derivation of the
factorizing twist on a generic evaluation representation. We find
that the triangular part of the twist is determined by the requirement
that the coproduct of one of the algebra generators (here $\Delta
D(u)$) in this representation be diagonal after twisting. This is
based on the same idea as the modified Functional Bethe Ansatz
in~\cite{Te99}.

The diagonal part of the twist is discussed in
Section~\ref{sect_diagonal}. We find that the diagonal part can be
determined by the additional requirement that the twisted coproduct be
cocommutative. The result comes as a set of recursion relations for
the coefficients of the diagonal part which can be solved in such a
way that the resulting form of the diagonal part is independent of the
particular representations.

In Section~\ref{sect_universal} we demonstrate in detail how the twist
factorizes the universal $R$-matrix of $Y(\ssl_2)$ on evaluation
representations. We summarize how the different ways of twisting which
are based on the diagonalization of different generators or on the use
of different diagonal parts for the twist, are related. Finally
we determine the precise conditions for the existence of the
factorizing twist on evaluation representations.

Section~\ref{sect_conclusion} contains a summary and comments on
future directions of research and important open questions.

%
\section{Preliminaries}
%
\label{sect_prep}

\subsection{Factorizing twists}

In order to fix the notation let us first recall the key definitions
and theorems about factorizing twists. They are due to
Drinfel'd~\cite{Dr87,Dr90} where more details can be found. We write
$\mu,\eta,\Delta,\epsilon,S$ for the product, unit, coproduct, co-unit
resp.~antipode of a Hopf algebra.

\begin{definition}
A Hopf algebra $\sym{A}$ is called \emph{quasi-triangular} if there
exists an invertible element $R\in\sym{A}\otimes\sym{A}$, called the
\emph{universal $R$-matrix}, which satisfies
\begin{mathletters}
\begin{eqnarray}
\label{eq_nonlineara}
  (\Delta\otimes\id)(R) &=& R_{13}R_{23},\\
\label{eq_nonlinearb}
  (\id\otimes\Delta)(R) &=& R_{13}R_{12},\\
\label{eq_almostcoc}
  \Dop(a) &=& R\cdot\Delta(a)\cdot R^{-1},
\end{eqnarray}%
\end{mathletters}%
for all $a\in\sym{A}$. It is called \emph{triangular} if in
addition
\begin{equation}
  \alignidx{R_{21}=R_{12}^{-1}}.
\end{equation}
Here $R_{ij}$ denote as usual the actions of $R$ on the different
factors of $\sym{A}\otimes\sym{A}\otimes\sym{A}$.
\end{definition}

\begin{definition}
Let $\sym{A}$ be a Hopf algebra. An invertible element
$F\in\sym{A}\otimes\sym{A}$ is called a \emph{co-unital $2$-cocycle} or
\emph{Drinfel'd twist} if it satisfies
\begin{mathletters}
\begin{eqnarray}
  (\epsilon\otimes\id)(F) &=& 1,\\
  (\id\otimes\epsilon)(F) &=& 1,\\
\label{eq_cocycle}
  F_{12}\cdot (\Delta\otimes\id)(F) &=& F_{23}\cdot (\id\otimes\Delta)(F).
\end{eqnarray}%
\end{mathletters}%
\end{definition}

\begin{theorem}[Drinfel'd]
\label{thm_twist}
Let $\sym{A}$ be a quasi-triangular Hopf algebra and
$F\in\sym{A}\otimes\sym{A}$ be a co-unital $2$-cocycle. Then the
algebra of $\sym{A}$ together with the operations
\begin{mathletters}
\begin{eqnarray}
  \Delta_F(a) &:=& F\cdot\Delta(a)\cdot F^{-1},\\
\label{eq_antipode}
  S_F(a) &:=& u\cdot S(a)\cdot u^{-1},\qquad u:= \mu(\id\otimes S)(F),\\
  R_F &:=& F_{21}\cdot R\cdot F^{-1}
\end{eqnarray}%
\end{mathletters}%
and the old co-unit $\epsilon$, forms a quasi-triangular Hopf algebra
$\sym{A}_F$. 
\end{theorem}

The cocycle condition~\eqref{eq_cocycle} is required in order to make
the twisted coproduct $\Delta_F$ co-associative so that one obtains a
Hopf algebra rather than just a quasi-Hopf algebra.

\begin{definition}
Let $\sym{A}$ be a quasi-triangular Hopf algebra with a co-unital
$2$-cocycle $F\in\sym{A}\otimes\sym{A}$. $F$ is called a
\emph{factorizing twist} if $R_F=1\otimes 1$ in
Theorem~\ref{thm_twist}, \ie\
\begin{equation}
  \alignidx{R_{12}=F_{21}^{-1}\cdot F_{12}}.
\end{equation}
\end{definition}

\begin{remark}
\label{rem_opposite}
\begin{myenumerate}
\item
  If a quasi-triangular Hopf algebra $\sym{A}$ admits a factorizing
  twist, the twisted coproduct $\Delta_F$ is cocommutative.
\item
  In this case the Hopf algebra is triangular since its universal
  $R$-matrix satisfies
\begin{equation}
  \alignidx{R_{21}=F_{12}^{-1}\cdot F_{21}=R_{12}^{-1}}.
\end{equation}
\item
  The converse implication is true at least for finite-dimensional
  semi-simple Hopf algebras~\cite{EtGe98a}. Up to complications due to
  the fact that the Yangian $Y(\ssl_2)$ has only a pseudo-universal
  $R$-matrix, the results in~\cite{Te99} and in this paper
  suggest that it applies to $Y(\ssl_2)$ as well.
\item
  If $F$ is a factorizing twist, then the opposite coproduct can be
  made cocommutative using the twist $F_{21}$ rather than $F_{12}$:
\begin{equation}
  F_{21}\cdot\Dop X\cdot F_{21}^{-1}
    = F_{12}\cdot\Delta X\cdot F_{12}^{-1}.
\end{equation}
\item
  Let $\gamma\in\sym{A}$ be invertible and $\epsilon(\gamma)=1$, then
\begin{equation}
  F_\gamma = (\gamma\otimes\gamma)\cdot F\cdot{(\Delta\gamma)}^{-1}
\end{equation}
  is another co-unital $2$-cocycle. Furthermore the Hopf algebras
  $\sym{A}_F$ and $\sym{A}_{F_\gamma}$ obtained by twisting with $F$
  resp.\ $F_\gamma$ are isomorphic under an inner automorphism. The
  twists $F$ and $F_\gamma$ are called cohomologous in this case.
  These `gauge' degrees of freedom can be understood in terms of
  non-Abelian cohomology, see \eg~\cite{Ma95a}.
\end{myenumerate}
\end{remark}

\subsection{Notations and conventions}

In this section we explain our notation for representations of the Lie
algebra $\ssl_2$ and for evaluation representations of the Yangian
$Y(\ssl_2)$. 

\subsubsection{Representations of the $\ssl_2$ Lie algebra}

We choose a Cartan-Weyl basis $(H,E,F)$ for the Lie algebra $\ssl_2$,
\begin{equation}
[H,E]=E,\qquad [H,F]=-F,\qquad [E,F]=2 H.
\end{equation}
Let $\lambda$ denote the highest weight of the representation
$V_\lambda$, $\ket{0}$ be the highest weight vector and
$(\ket{0},\ket{1},\ldots)$ a weight basis such that
\begin{equation}
 H\ket{k}=(\lambda-k)\ket{k},\quad k\in\N_0,
\end{equation}
\ie\ we have finite-dimensional representations of dimension
$2\,\lambda+1$ for $2\,\lambda\in\N_0$. A representation for
which the eigenvalue of the Casimir element
$C^{(2)}=H^2+\frac{1}{2}(EF+FE)$ equals $\lambda(\lambda+1)$, is
given by
\begin{mathletters}
\begin{eqnarray}
 E\ket{k}&=&e_k\ket{k-1},\\
 F\ket{k}&=&f_k\ket{k+1},
\end{eqnarray}%
\end{mathletters}%
where $e_k=k$ and $f_k=2\,\lambda-k$.

In the following, we use the same notation for finite-dimensional and
infinite-dimensional highest weight representations. This means that
in the finite-dimensional case the sequence of basis vectors
$(\ket{0},\ket{1},\ldots)$ terminates, and \eg\ in the case of
dimension $2\,\lambda+1$, $2\,\lambda\in\N_0$, the vectors $\ket{k}$
for $k>2\,\lambda$ can be omitted or set to zero in all formulas.

Likewise, for the tensor product $V_{\lambda_1}\otimes V_{\lambda_2}$,
the subspace of weight $\lambda_1+\lambda_2-m$ is spanned by
$(\ket{m}\otimes\ket{0},\ket{m-1}\otimes\ket{1},\ldots,\ket{0}\otimes\ket{m})$.
In the finite-dimensional case we have $\ket{\ell}\otimes\ket{k}=0$ if
$\ell>2\,\lambda_1$ or $k>2\,\lambda_2$, \ie\ a number of vectors can
be omitted from the generating set. In the following, we arrange all
statements in such a way that this does not change the formulas.

\subsubsection{The Yangian $Y(\ssl_2)$}

The Yangian $Y(\ssl_2)$ is the free associative algebra over $\C$
generated by $t_{ij}^{(r)}$, $i,j\in\{1,2\}$, $r\in\N_0$, modulo
the ideal generated by the relations
\begin{gather}
  R^{ij}_{k\ell}(u-v)\,T^k_m(u)\, T^\ell_n(v)
    = T^j_\ell(v)\,T^i_k(u)\,R^{k\ell}_{mn}(u-v),\\
  \qdet T(u)=1.
\end{gather}
for all $i,j,m,n\in\{1,2\}$, where summation over $k$ and $\ell$ is
understood. Here 
$R^{ij}_{k\ell}(u)=\delta^i_\ell\delta^j_k+\delta^i_k\delta^j_\ell\cdot
1/u$, and the power series
\begin{equation}
\label{eq_functionals}
  T^i_j(u)=\delta_{ij} + \sum_{r=1}^\infty t_{ij}^{(r)}\,u^{-r}
\end{equation}
are to be expanded. The quantum determinant is defined as
\begin{equation}
\label{eq_qdet}
  \qdet T(u)=T^1_1\bigl(u+\frac{1}{2}\bigr)\,T^2_2\bigl(u-\frac{1}{2}\bigr)
  - T^1_2\bigl(u+\frac{1}{2}\bigr)\,T^2_1\bigl(u-\frac{1}{2}\bigr).
\end{equation}

\subsubsection{Evaluation representations of $Y(\ssl_2)$}

For a given representation $V_\lambda$ of the Lie algebra $\ssl_2$, we
obtain an evaluation representation $V_\lambda(\delta)$ of $Y(\ssl_2)$
setting
\begin{equation}
\label{eq_evaluation}
  \begin{aligned}
    A(u)&=T^1_1(u)=u-\delta+\eta\,H,\\
    C(u)&=T^2_1(u)=\eta\,E,
  \end{aligned}\qquad\begin{aligned} 
    B(u)&=T^1_2(u)=\eta\,F\\
    D(u)&=T^2_2(u)=u-\delta-\eta\,H,
  \end{aligned}
\end{equation}
for the functionals in~\eqref{eq_functionals}. Here the parameter
$\delta$ is associated with the representation, $\eta$ is a constant
which could be eliminated by a rescaling, but is part of the common
notation. In particular~\eqref{eq_evaluation} involves a homomorphism
of algebras $Y(\ssl_2)\to U(\ssl_2)$. In~\eqref{eq_evaluation} we have
chosen a normalization different from the usual one. This will make
the calculations easier, but will not affect the structure of the
results. The quantum determinant, however, is now given by
\begin{equation}
  \qdet T(u)=\Bigl(u+\frac{\eta}{2}\Bigr)\,\Bigl(u-\frac{\eta}{2}\Bigr)
  - \eta^2\,\lambda(\lambda+1),
\end{equation}
so strictly speaking we make calculations in $V_\lambda(\delta)$ as in
a representation of $Y(\gl_2)$. For the action of the operators
defined in~\eqref{eq_evaluation} on $V_{\lambda_j}(\delta_j)$ we write
\begin{mathletters}
\label{eq_defineabbrev}
\begin{alignat}{2}
A(u)\ket{k} &= (u-\delta_j+\eta\,(\lambda_j-k))\ket{k} &&=: a^{(j)}_k(u)\ket{k},\\
B(u)\ket{k} &= \eta\,F\ket{k}                          &&=: \eta\,f^{(j)}_k\ket{k+1},\\
C(u)\ket{k} &= \eta\,E\ket{k}                          &&=: \eta\,e^{(j)}_k\ket{k-1},\\
D(u)\ket{k} &= (u-\delta_j-\eta\,(\lambda_j-k))\ket{k} &&=: d^{(j)}_k(u)\ket{k},
\end{alignat}%
\end{mathletters}%
where $e^{(j)}_kf^{(j)}_{k-1}=k\,(2\,\lambda_j-k+1)$. 

The coproduct in $Y(\ssl_2)$ can be written
\begin{equation}
  \begin{aligned}
    \Delta A(u)&=A(u)\otimes A(u)+C(u)\otimes B(u),\\
    \Delta C(u)&=A(u)\otimes C(u)+C(u)\otimes D(u),
  \end{aligned}\quad
  \begin{aligned}
    \Delta B(u)&=B(u)\otimes A(u)+D(u)\otimes B(u),\\
    \Delta D(u)&=B(u)\otimes C(u)+D(u)\otimes D(u).
  \end{aligned}
\end{equation}
It might seem more natural to call this $\Dop$ rather than $\Delta$
since it is opposite compared with the usual matrix product, but our
results look more natural if this coproduct is used. It is the same as
defined in~(2.14) of~\cite{Te99}.

The coproduct then acts on $V_{\lambda_1}(\delta_1)\otimes
V_{\lambda_2}(\delta_2)$ as
\begin{mathletters}
\label{eq_deltad}
\begin{alignat}{2}
  \Delta A(u)\ket{\ell,k} &= a_\ell^{(1)}(u)\,a_k^{(2)}(u)\,\ket{\ell,k} 
    &&+ \eta^2e_\ell^{(1)}f_k^{(2)}\ket{\ell-1,k+1},\\
  \Delta B(u)\ket{\ell,k} &= \eta\,f_\ell^{(1)}a_k^{(2)}(u)\,\ket{\ell+1,k}
    &&+ \eta\,d_\ell^{(1)}(u)\,f_k^{(2)}\ket{\ell,k+1},\\
  \Delta C(u)\ket{\ell,k} &= \eta\,a_\ell^{(1)}(u)\,e_k^{(2)}\ket{\ell,k-1}
    &&+ \eta\,e_\ell^{(1)}d_k^{(2)}(u)\,\ket{\ell-1,k},\\
  \Delta D(u)\ket{\ell,k} &= \eta^2f_\ell^{(1)}e_k^{(2)}\ket{\ell+1,k-1}
    &&+ d_\ell^{(1)}(u)\,d_k^{(2)}(u)\,\ket{\ell,k},
\end{alignat}%
\end{mathletters}%
where $\ket{\ell,k}:=\ket{\ell}\otimes\ket{k}$.

\subsection{Irreducibility of evaluation representations}

When we study the conditions for the existence of the factorizing
twists, we refer to the corresponding conditions for the existence of
$R$-matrices. Here we state a few results which are due to Tarasov and
can be found \eg\ in~\cite{ChPr90,NaTa98}:

\begin{theorem}[Tarasov]
Each finite-dimensional irreducible representation of the Yangian
$Y(\ssl_2)$ over $\C$ is isomorphic to a tensor product of evaluation
representations. Two such tensor products describe isomorphic
representations if and only if they are related by a permutation of
the tensor factors.
\end{theorem}

\begin{theorem}[Tarasov]
\label{thm_irreducible}
The tensor product of finite-dimensional evaluation representations
$V_{\lambda_1}(\delta_1)\otimes V_{\lambda_2}(\delta_2)$ is reducible
if and only if
\begin{equation}
\label{eq_reducible}
  \pm(\delta_1-\delta_2)/\eta = \lambda_1+\lambda_2 - j + 1
\end{equation}  
for an integer $j$ satisfying
$0<j\leq\min\{2\,\lambda_1,2\,\lambda_2\}$. In this case the
representation is not completely reducible nor isomorphic to
$V_{\lambda_2}(\delta_2)\otimes V_{\lambda_1}(\delta_1)$.
\end{theorem}

\begin{remark}
The existence of these representations is closely related to the
failure of the Yangian to be quasi-triangular. In particular the
condition~\eqref{eq_almostcoc} is not satisfied in this case.  In
evaluation representations the $R$-matrix is given by a matrix
$R(\delta_1-\delta_2)$ of rational functions of
$\delta_1-\delta_2$. This matrix is fixed by~\eqref{eq_almostcoc}
which reads here
\begin{equation}
  \Dop X(u) = 
    R(\delta_1-\delta_2)\cdot\Delta X(u)\cdot {R(\delta_1-\delta_2)}^{-1},
\end{equation}
but has got a pole or fails to be invertible at the
`singularities'~\eqref{eq_reducible}. The simplest example is the case
$V_{1/2}(\delta_1)\otimes V_{1/2}(\delta_2)$ where
\begin{equation}
  R(\delta_1-\delta_2)=\frac{\delta_1-\delta_2}{\delta_1-\delta_2+\eta}\,\openone
  + \frac{\eta}{\delta_1-\delta_2+\eta}\,P,
\end{equation}
where $P(a\otimes b)=b\otimes a$. In the case of $\delta_1-\delta_2=\pm\eta$,
either $R$ or $R^{-1}$ do not exist.
\end{remark}

%
\section{The triangular part of the factorizing twist}
%
\label{sect_triangular}

The essence of the modification of the Functional Bethe Ansatz used
in~\cite{Te99} in order to construct factorizing twists is the
diagonalization of the operator $\Delta D(u)$ on
$V_{\lambda_1}(\delta_1)\otimes V_{\lambda_2}(\delta_2)$ (one could
equally well have used $\Delta A(u)$).

In the following, we will not use a modified Functional Bethe Ansatz,
but rather perform the necessary change of basis explicitly. We find that
the prescription to diagonalize $\Delta D(u)$ fixes the
triangular part of the factorizing twist. Diagonalizing $\Delta A(u)$
likewise results in the triangular part of (another) factorizing
twist. However, it is not possible to diagonalize both $\Delta
A(u)$ and $\Delta D(u)$ simultaneously in a way that is independent of $u$.

\subsection{Eigenvectors of some special matrices}

A few technical results are needed to accomplish the
diagonalization. The following statements are all elementary and
straight-forward to prove.

\begin{lemma}
\label{lemma_diagonalize}
The eigenvectors of the $(n+1)\times(n+1)$ real or complex matrix
\begin{equation}
\label{eq_matrix_a}
  M=\begin{pmatrix}
      c_0&b_0&      &       &\\ 
         &c_1&b_1   &       &\\ 
         &   &\ddots&\ddots &\\
         &   &      &c_{n-1}&b_{n-1}\\ 
         &   &      &       &c_n
     \end{pmatrix},
\end{equation}
with pairwise distinct $c_i$ are the column vectors $v^{(k)}$, $0\leq
k\leq n$, with the components 
\begin{equation}
  v_j^{(k)}=\begin{cases}
              \quad 0,&\text{if}\quad j>k,\\ 
              \quad \displaystyle\prod_{i=j}^{k-1}\frac{b_i}{c_k-c_i},&\text{if}\quad j\leq k,
            \end{cases}
\end{equation}
$0\leq j\leq n$, where the empty product is by definition equal to
$1$. The eigenvalue of $v^{(k)}$ is $c_k$. 
\end{lemma}

\begin{corollary}
The matrix $M$ in~\eqref{eq_matrix_a} can be diagonalized using
\begin{equation}
  S^{-1}\cdot M\cdot S=\diag (c_0,\ldots,c_n),
\end{equation}
where the change of basis is given by the (upper triangular) matrix
with coefficients
\begin{equation}
\label{eq_matrix_s}
  S_{jk}=v_j^{(k)},
\end{equation}
$0\leq j,k\leq n$, where the first index denotes the row and the
second the column of the matrix.
\end{corollary}

\begin{lemma}
For pairwise distinct $c_1,\ldots,c_m$ the following sum vanishes:
\begin{equation}
  \sum_{i=1}^m\prod_{\substack{j=1\\ j\neq i}}^m\frac{1}{c_j-c_i}=0.
\end{equation}
\end{lemma}
The proof is again elementary. An elegant argument can be found \eg\
in~\cite{HoMa99}. This identity is necessary to find the inverse matrix:

\begin{corollary}
The inverse matrix of $S$ in~\eqref{eq_matrix_s} is given by
\begin{equation}
  S^{-1}_{jk}=\begin{cases}
                 \quad 0,&\text{if}\quad j>k,\\
                 \quad \displaystyle\prod_{i=j+1}^k\frac{b_{i-1}}{c_j-c_i},&\text{if}\quad j\leq k,
              \end{cases}
\end{equation}
$0\leq j,k\leq n$. It is again upper triangular.
\end{corollary}

We will also need analogous statements referring to the transposed matrix:
\begin{corollary}
\label{cor_diagtransposed}
The transposed matrix $\tilde M=M^\tp$ is diagonalized by
\begin{equation}
\tilde S^{-1}\cdot \tilde M\cdot\tilde S=\diag(c_0,\ldots,c_n),
\end{equation}
where $\tilde S={(S^{-1})}^\tp$ is now lower triangular.
\end{corollary}

\subsection{Diagonalizing $\Delta D(u)$}

The matrix of $\Delta D(u)$ on $V_{\lambda_1}(\delta_1)\otimes
V_{\lambda_2}(\delta_2)$ decomposes into blocks acting on invariant
subspaces $V_m$ which are spanned by
\begin{equation}
  V_m=\Span\{\ket{m,0},\ket{m-1,1},\ldots,\ket{0,m}\},
\end{equation}
$m\in\N_0$. These subspaces are always finite-dimensional even if the
representations under study are infinite-dimensional. In that case
there are just infinitely many blocks of growing but finite size.

On each $V_m$, $\Delta D(u)$ has the form
\begin{equation}
\Delta D(u)=\begin{pmatrix}
             c_0&   &      &       &\\ 
             b_0&c_1&      &       &\\ 
                &b_1&\ddots&       &\\
                &   &\ddots&c_{m-1}&\\ 
                &   &      &b_{m-1}&c_m
           \end{pmatrix},
\end{equation}
where the matrix coefficients belong to
$\ket{0,m},\ket{1,m-1},\ldots,\ket{m,0}$ from top to bottom and left
to right. We find using~\eqref{eq_deltad}
\begin{mathletters}
\begin{eqnarray}
  c_k        &=& d_k^{(1)}(u)\cdot d_{m-k}^{(2)}(u),\\
  b_k        &=& \eta^2f_k^{(1)}e_{m-k}^{(2)},\\
\label{eq_useg}
  c_\ell-c_k &=& \eta\,(\ell-k)\,g(m-k-\ell),
\end{eqnarray}%
\end{mathletters}%
where $g(x):=\delta_1-\delta_2+\eta(\lambda_1-\lambda_2+x)$.
Therefore each of the blocks of $\Delta D(u)$ can be diagonalized using
Corollary~\ref{cor_diagtransposed}.

\begin{proposition}
\label{prop_eigenvectd}
We label the eigenvectors $v_{\ell k}$ of $\Delta D(u)$ by two
indices, $k,\ell\in\N_0$. These eigenvectors are given by
\begin{equation}
\label{eq_eigenvectd}
\begin{split}
  v_{\ell k}&=q_{\ell k}\sum_{j=0}^{k+\ell}\tilde S_{j\ell}\ket{j,k+\ell-j}\\
            &=q_{\ell k}\sum_{n=0}^k\prod_{i=1}^n
                       \frac{b_{\ell+i-1}}{c_\ell-c_{j+\ell}}
                       \ket{\ell+n,k-n}\\
            &=q_{\ell k}\sum_{n=0}^kM^{k,\ell}_n\ket{\ell+n,k-n},
\end{split}
\end{equation}
for some coefficients $q_{\ell k}$. We have defined
\begin{equation}
\label{eq_definem}
  M^{k,\ell}_n := \frac{{(-\eta)}^n}{n!}\,\prod_{j=1}^n
                   \frac{f_{\ell+j-1}^{(1)}e_{k-j+1}^{(2)}}{g(k-\ell-j)}.
\end{equation}
The eigenvalue of $v_{\ell k}$ is $d_\ell^{(1)}(u)\cdot
d_k^{(2)}(u)$. In particular the eigenvectors do not depend on $u$.
\end{proposition}

Here the coefficients $q_{\ell k}$ reflect the freedom of normalizing
the eigenvectors. They will be fixed using other conditions in
Section~\ref{sect_diagonal}. It is amazing that this change of basis
already contains the full information about one of the two triangular
parts of the Gauss decomposition of the the $R$-matrix on a generic
evaluation representation. This fact will become clear in
Section~\ref{sect_universal}. Similarly we find

\begin{proposition}
The inverse transformation of~\eqref{eq_eigenvectd} is given by
\begin{equation}
\label{eq_eigenvectdi}
  \ket{\ell,k} = \sum_{n=0}^k\frac{\eta^n}{n!}\prod_{j=1}^n
    \frac{f_{\ell+j-1}^{(1)}e_{k-j+1}^{(2)}}{g\bigl((k-n)-(\ell+n)+j\bigr)}\,
    \alignidx{q_{\ell+n,k-n}^{-1}v_{\ell+n,k-n}}.
\end{equation}
\end{proposition}
We will need the following properties of the coefficients
$M^{k,\ell}_n$ later:

\begin{lemma}
\label{lemma_propertym}
The matrix elements $M^{k,\ell}_n$ defined in~\eqref{eq_definem} satisfy
\begin{mathletters}
\begin{eqnarray}
  M^{k,\ell}_0&=&1,\\
  \frac{M^{k,\ell}_{n+1}}{M^{k,\ell}_n} 
    &=& -\frac{\eta\,f_{\ell+n}^{(1)}e_{k-n}^{(2)}}{(n+1)\,g(k-\ell-n-1)},\\
  \frac{M^{k,\ell-1}_n}{M^{k,\ell}_n}
    &=& \frac{f_{\ell-1}^{(1)}g(k-\ell-n)}{f_{\ell+n-1}^{(1)}g(k-\ell)},\\
  \frac{M^{k-1,\ell}_n}{M^{k,\ell}_n}
    &=& \frac{e_{k-n}^{(2)}g(k-\ell-1)}{e_{k}^{(2)}g(k-\ell-n-1)},\\
\end{eqnarray}%
\end{mathletters}%
\end{lemma}

\subsection{Diagonalizing $\Delta A(u)$}

Likewise, it is possible to diagonalize $\Delta A(u)$. Again its
matrix decomposes into blocks on the subspaces $V_m$ which are of the form
\begin{equation}
  \Delta A(u)=\begin{pmatrix}
               c_0&b_0&      &       &\\ 
                  &c_1&b_1   &       &\\ 
                  &   &\ddots&\ddots &\\
                  &   &      &c_{m-1}&b_{m-1}\\
                  &   &      &       &c_m
            \end{pmatrix}.
\end{equation}
Here
\begin{mathletters}
\begin{eqnarray}
  c_k        &=& a_k^{(1)}(u)\cdot a_{m-k}^{(2)}(u),\\
  b_k        &=& \eta^2e_{k+1}^{(1)}f_{m-k-1}^{(2)},\\
  c_\ell-c_k &=& \eta\,(\ell-k)\,\tilde g(m-k-\ell),
\end{eqnarray}%
\end{mathletters}%
where now $\tilde g(x):=\delta_2-\delta_1+\eta\,(\lambda_1-\lambda_2+x)$,
\ie\ compared with $g(x)$ in~\eqref{eq_useg} we just exchange
$\delta_1\leftrightarrow\delta_2$. 

Each of the blocks of $\Delta A(u)$ can thus be diagonalized using
Lemma~\ref{lemma_diagonalize}. 

\begin{proposition}
The eigenvectors of $\Delta A(u)$ on $V_{\lambda_1}(\delta_2)\otimes
V_{\lambda_2}(\delta_2)$ are given by $\tilde v_{\ell k}$,
$k,\ell\in\N_0$, 
\begin{equation}
\label{eq_eigenvecta}
\begin{split}
  \tilde v_{\ell k} &= \tilde q_{\ell k}\sum_{j=0}^{k+\ell}S_{j\ell}\ket{j,k+\ell-j}\\
                    &= \tilde q_{\ell k}\sum_{n=0}^\ell \tilde M^{k,\ell}_n\ket{\ell-n,k+n},
\end{split}
\end{equation}
where we have abbreviated
\begin{equation}
\label{eq_definemtilde}
  \tilde M^{k,\ell}_n := \frac{\eta^n}{n!}\prod_{j=1}^n
    \frac{e_{\ell-j+1}^{(1)}f_{k+j-1}^{(2)}}{\tilde g(k-\ell+j)}.
\end{equation}
The eigenvalue of $\tilde v_{\ell k}$ is $a_\ell^{(1)}(u)\cdot a_k^{(2)}(u)$.
\end{proposition}
Similarly we find

\begin{proposition}
The inverse transformation of~\eqref{eq_eigenvecta} is given by
\begin{equation}
\label{eq_eigenvectai}
  \ket{\ell,k} = \sum_{n=0}^\ell\frac{{(-\eta)}^{n}}{n!}\prod_{j=1}^n
    \frac{e_{\ell-j+1}^{(1)}f_{k+j-1}^{(2)}}{\tilde g\bigl((k+n)-(\ell-n)-j\bigr)}\,
    \alignidx{\tilde q_{\ell-n,k+n}^{-1}\,\tilde v_{\ell-n,k+n}}.
\end{equation}
\end{proposition}
The following properties of the coefficients $\tilde M^{k,\ell}_n$
will later be useful:

\begin{lemma}
\label{lemma_propertymtilde}
The matrix elements $\tilde M^{k,\ell}_n$ defined
in~\eqref{eq_definemtilde} satisfy:
\begin{mathletters}
\begin{eqnarray}
  \tilde M^{k,\ell}_0 &=& 1,\\
  \frac{\tilde M^{k,\ell}_{n+1}}{\tilde M^{k,\ell}_n}
    &=& \frac{\eta\,e_{\ell-n}^{(1)}f_{k+n}^{(2)}}{(n+1)\,\tilde g(k-\ell+n+1)},\\
  \frac{\tilde M^{k,\ell-1}_n}{\tilde M^{k,\ell}_n}
    &=& \frac{e_{\ell-n}^{(1)}\,\tilde g(k-\ell+1)}{e_\ell^{(1)}\,\tilde g(k-\ell+n+1)},\\
  \frac{\tilde M^{k-1,\ell}_n}{\tilde M^{k,\ell}_n}
    &=& \frac{f_{k-1}^{(2)}\,\tilde g(k-\ell+n)}{f_{k+n-1}^{(2)}\,\tilde g(k-\ell)},\\
\end{eqnarray}%
\end{mathletters}%
\end{lemma}

\subsection{Towards a universal expression}
\label{sect_univ}

In order to derive a form of the change of basis~\eqref{eq_eigenvectd}
diagonalizing $\Delta D(u)$ which does not depend on the particular
representation, it is necessary to express the basis vectors
$\ket{\ell+n,k-n}$ in terms of $\ket{\ell,k}$. Using
\begin{equation}
  F^n\otimes E^n\ket{\ell,k} 
    = \biggl(\prod_{j=1}^n f_{\ell+j-1}^{(1)}e_{k-j+1}^{(2)}\biggr)\ket{\ell+n,k-n}
\end{equation}
and
\begin{equation}
  \Ha\ket{\ell,k} = (\lambda_1-\ell)\ket{\ell,k},\qquad
  \Hb\ket{\ell,k} = (\lambda_2-k)\ket{\ell,k},
\end{equation}
equation~\eqref{eq_eigenvectd} reads
\begin{equation}
\label{eq_diagonalized}
  v_{\ell k} = q_{\ell k}\sum_{n=0}^k\frac{{(-\eta)}^n}{n!}F^n\otimes E^n
  \Bigl(\prod_{j=1}^n{(\delta_1-\delta_2+\eta\,(\Ha-\Hb)-\eta\,j)}^{-1}\Bigr)
  \ket{\ell,k}.
\end{equation}
It is apparent that in order to generalize this to generic highest
weight representations, it is sufficient to make the sum infinite. We
conclude: 

\begin{proposition}
The expression
\begin{equation}
\label{eq_f12i}
  F_{12}^{-1} = \biggl(\sum_{n=0}^\infty\frac{{(-\eta)}^n}{n!}F^n\otimes E^n\prod_{j=1}^n
                {(\delta_1-\delta_2+\eta\,(\Ha-\Hb)-\eta\,j)}^{-1}\biggr)\,Q_{12}^{-1},
\end{equation}
where we denote
\begin{equation}
  Q_{12}^{-1}\ket{\ell,k} = q_{\ell k}\ket{\ell,k},
\end{equation}
specializes to the change of basis~\eqref{eq_diagonalized}
and~\eqref{eq_eigenvectd} on all representations
$V_{\lambda_1}(\delta_1)\otimes V_{\lambda_2}(\delta_2)$. Recall that
this change of basis diagonalizes $\Delta D(u)$.
\end{proposition}

\begin{remark}
\begin{myenumerate}
\item
  The result~\eqref{eq_f12i} agrees with~(4.53) in~\cite{Te99}.
\item
  The operator $\alignidx{F_{12}^{-1}\cdot Q_{12}}$ decomposes into (possibly
  infinitely many) triangular matrices on each
  representation. Its diagonal elements are given by the $n=0$
  summand, \ie\ all its coefficients on the diagonal equal $1$. This
  is the triangular part of the Gauss decomposition of the twist while
  $Q_{12}^{-1}$ describes its diagonal part.
\item
  The question under which conditions the expression~\eqref{eq_f12i}
  is well-defined is discussed in Section~\ref{sect_existence}.
\end{myenumerate}
\end{remark}
Similarly, we can obtain the inverse transformation starting
with~\eqref{eq_eigenvectdi}.

\begin{proposition}
The expression
\begin{equation}
\label{eq_f12}
  F_{12} = Q_{12}\sum_{n=0}^\infty\frac{\eta^n}{n!}\biggl(\prod_{j=1}^n
           {(\delta_1-\delta_2+\eta\,(\Ha-\Hb)+\eta\,j)}^{-1}\biggr)F^n\otimes E^n.
\end{equation}
specializes to the inverse change of basis~\eqref{eq_eigenvectdi} for
the diagonalization of $\Delta D(u)$ on all representations
$V_{\lambda_1}(\delta_1)\otimes V_{\lambda_2}(\delta_2)$.
\end{proposition}

Analogous results can be obtained for the case in which $\Delta A(u)$
rather than $\Delta D(u)$ is diagonalized. 

\begin{proposition}
The expressions
\begin{eqnarray}
  \tilde F_{12}^{-1}&=&\biggl(\sum_{n=0}^\infty\frac{\eta^n}{n!}E^n\otimes F^n\prod_{j=1}^n
                {(\delta_2-\delta_1+\eta\,(\Ha-\Hb)+\eta\,j)}^{-1}\biggr)\,\tilde Q_{12}^{-1},\\
  \tilde F_{12} &=& \tilde Q_{12}\sum_{n=0}^\infty\frac{{(-\eta)}^n}{n!}\biggl(\prod_{j=1}^n
           {(\delta_2-\delta_1+\eta\,(\Ha-\Hb)-\eta\,j)}^{-1}\biggr)E^n\otimes F^n,
\end{eqnarray}
specialize to the change of basis~\eqref{eq_eigenvecta} resp.\ its
inverse transformation~\eqref{eq_eigenvectai} on all representations
$V_{\lambda_1}(\delta_1)\otimes V_{\lambda_2}(\delta_2)$. Recall that
this change of basis diagonalizes $\Delta A(u)$. Here we write
\begin{equation}
  \tilde Q_{12}^{-1}\ket{\ell,k} = \tilde q_{\ell k}\ket{\ell,k}.
\end{equation}
\end{proposition}

%
\section{The diagonal part of the factorizing twist}
%
\label{sect_diagonal}

In this section we study the conditions that restrict the
$q_{\ell k}$ coefficients of the diagonal part of the twist.

\subsection{Cocommutativity}

In order to make $F_{12}$ in~\eqref{eq_f12i} a factorizing twist,
it is of course not sufficient to diagonalize $\Delta D(u)$. The
desired property is that the twisted Yangian coproduct 
$F\cdot \Delta X(u)\cdot F^{-1}$ be cocommutative. It is
thus necessary to fix the coefficients $q_{\ell k}$ of $Q_{12}^{-1}$
appropriately.

A case by case analysis of the action of \eg\ $\Delta B(u)$ in the new
basis $v_{\ell k}$, see~\eqref{eq_diagonalized}, in special evaluation
representations shows that $F\cdot\Delta B(u)\cdot F^{-1}$ is of the
form
\begin{equation}
  \Delta B(u)v_{\ell k} 
    = \alpha_{\ell k}(u)\cdot\frac{q_{\ell k}}{q_{\ell+1,k}}\,v_{\ell+1,k}
    + \beta_{\ell k}(u)\cdot\frac{q_{\ell k}}{q_{\ell,k+1}}\,v_{\ell,k+1},
\end{equation}
where $\alpha_{\ell k}(u), \beta_{\ell k}(u)$ are as yet unspecified
functions of $u$. Cocommutativity means that the expression for
$F\cdot\Delta B(u)\cdot F^{-1}$ written in $U(\ssl_2)$ is symmetric,
\ie\ that
\begin{equation}
  \alpha_{\ell k}(u)\cdot\frac{q_{\ell k}}{q_{\ell+1,k}}
  = {\biggl(\beta_{k\ell}(u)\cdot\frac{q_{k\ell}}{q_{k+1,\ell}}
      \biggr)}_{1\leftrightarrow 2},
\end{equation}
where $1\leftrightarrow 2$ indicates that
$\lambda_1\leftrightarrow\lambda_2$ and
$\delta_1\leftrightarrow\delta_2$ have to be exchanged in the
corresponding expressions.

It is apparent that there is a freedom to choose a multiplicative
factor in the coefficients $q_{\ell k}$ which depends only on $k+\ell$
and which is symmetric under the exchange $1\leftrightarrow 2$:
\begin{equation}
  q_{\ell k}\mapsto q_{\ell k}\cdot r_{k+\ell},\qquad
    r_{k+\ell}={(r_{k+\ell})}_{1\leftrightarrow 2}.
\end{equation}
Since we work only in representations, it is not easy to see whether
this freedom (which is also mentioned in~\cite{MaSa96,AlBo00})
exhausts the full freedom of choosing a cohomologous twist, see
Remark~\ref{rem_opposite}.

Finally it is possible to find a solution for the quotients $q_{\ell
k}/q_{\ell+1,k}$ \etc\ which makes the new coproduct $F\cdot \Delta
B(u)\cdot F^{-1}$ cocommutative. We show that in this case the
coproduct on the other generators is cocommutative as well. The
following proposition seems technically very complicated, but it is
just a reasonably straight-forward abstraction from explicit
calculations in particular finite-dimensional representations.

\begin{proposition}
\label{prop_newcoprod}
If the coefficients $q_{\ell k}$ of $Q_{12}^{-1}$ satisfy the recursion
relations
\begin{equation}
\label{eq_recursion}
  \frac{q_{\ell k}}{q_{\ell-1,k}} = \frac{g(-\ell)}{g(k-\ell)},\qquad
  \frac{q_{\ell,k+1}}{q_{\ell k}} = \frac{g(k-\ell)}{g(-2\,\lambda_1+k)},
\end{equation}
where $g(x)=\delta_1-\delta_2+\eta\,(\lambda_1-\lambda_2+x)$, then the
action of the coproducts on the basis vectors $v_{\ell k}$,
see~\eqref{eq_diagonalized}, is given by
\begin{eqnarray}
  \Delta D(u)\,v_{\ell k} &=& d_\ell^{(1)}(u)\cdot d_k^{(2)}(u)\,v_{\ell k},\\
  \Delta B(u)\,v_{\ell k} &=& 
    \eta f_\ell^{(1)}\,d_k^{(2)}(u)\,\frac{g(2\,\lambda_2-\ell)}{g(k-\ell)}v_{\ell+1,k}
  + \eta d_\ell^{(1)}(u)\,f_k^{(2)}\,\frac{g(-2\,\lambda_1+k)}{g(k-\ell)}v_{\ell,k+1},\\
  \Delta C(u)\,v_{\ell k} &=&
    \eta e_\ell^{(1)}\,d_k^{(2)}(u)\,\frac{g(-\ell)}{g(k-\ell)}v_{\ell-1,k}
  + \eta d_\ell^{(1)}(u)\,e_k^{(2)}\,\frac{g(k)}{g(k-\ell)}v_{\ell,k-1}.
\end{eqnarray}%
\end{proposition}

\begin{proof}
Diagonality of $\Delta D(u)$ was stated in
Proposition~\ref{prop_eigenvectd}. For $\Delta B(u)$ we first
calculate
\begin{equation}
\label{eq_dbnewbasis}
\begin{split}
  &\Delta B(u)\,v_{\ell k}\\
  &= q_{\ell k}\,\sum_{i=0}^kM^{k,\ell}_i\,\Delta B(u)\ket{\ell+i,k-i}\\
  &= q_{\ell k}\,\sum_{i=0}^kM^{k,\ell}_i\,\eta\,
        \bigl(d_{\ell+i}^{(1)}(u)f_{k-i}^{(2)}\ket{\ell+i,k-i+1}
         + f_{\ell+i}^{(1)}a_{k-i}^{(2)}(u)\ket{\ell+i+1,k-i}\bigr)\\
  &= q_{\ell k}\,\sum_{i=0}^kM^{k,\ell}_i\,\eta\,f_{\ell+i}^{(1)}\,\biggl(
         a_{k-i}^{(2)}(u)-d_{\ell+i+1}^{(1)}(u)\,
           \frac{\eta\,e_{k-i}^{(2)}f_{k-i-1}^{(2)}}{(i+1)g(k-\ell-i-1)}
         \biggr)\ket{\ell+i+1,k-i}\\
  &\qquad + q_{\ell k}\,\eta\,d_\ell^{(1)}(u)\,f_k^{(2)}\ket{\ell,k+1}.
\end{split}
\end{equation}
Here we use the notation of~\eqref{eq_defineabbrev}. In the third
line, the summation index was shifted and Lemma~\ref{lemma_propertym}
was used.

It remains to show that writing coefficients
\begin{equation}
  c^1_{\ell k} = \eta\,d_k^{(2)}(u)\,f_\ell^{(1)}\,\frac{g(2\,\lambda_2-\ell)}{g(k-\ell)},\qquad
  c^2_{\ell k} = \eta\,d_\ell^{(1)}(u)\,f_k^{(2)}\,\frac{g(-2\,\lambda_1+k)}{g(k-\ell)},
\end{equation}
the coproduct $\Delta B(u)$ has got the following form:
\begin{equation}
\label{eq_dbnewbasis2}
\begin{split}
  &\Delta B(u)\,v_{\ell k}\\
  &= c^1_{\ell k}v_{\ell+1,k} + c^2_{\ell k}v_{\ell,k+1}\\
  &= c^1_{\ell k}q_{\ell+1,k}\,\sum_{i=0}^kM^{k,\ell+1}_i\ket{\ell+i+1,k-i}
    + c^2_{\ell k}q_{\ell,k+1}\,\sum_{i=0}^{k+1}M^{k+1,\ell}_i\ket{\ell+i,k-i+1}\\
  &= \sum_{i=0}^kM^{k,\ell}_i\,\eta\,f^{(1)}_{\ell+i}\,\biggl(
      c^1_{\ell k}q_{\ell+1,k}\frac{g(k-\ell-1)}{\eta\,f_\ell^{(1)}\,g(k-\ell-i-1)}\\
  &\qquad - c^2_{\ell k}q_{\ell,k+1}\frac{e_{k+1}^{(2)}}{(i+1)\,g(k-\ell)}
      \biggr)\ket{\ell+i+1,k-i} + c^2_{\ell k}q_{\ell,k+1}\ket{\ell,k+1}.
\end{split}
\end{equation}
In the last line, the summation index was shifted and, using
Lemma~\ref{lemma_propertym} several times, the coefficient
$M^{k\,\ell}_i$ was cast in the same form as
in~\eqref{eq_dbnewbasis}. Both expressions~\eqref{eq_dbnewbasis}
and~\eqref{eq_dbnewbasis2} are equal if and only if for all
$k,\ell\in\N_0$
\begin{equation}
  q_{\ell k}\,\eta f_k^{(2)}d_\ell^{(1)}(u) = c^2_{\ell k}q_{\ell,k+1}
\end{equation}
and for all $i\in\{0,\ldots,k\}$
\begin{equation}
\begin{split}
  &a_{k-i}^{(2)}(u) - d_{\ell+i+1}^{(1)}(u)\,
    \frac{\eta\,e_{k-i}^{(2)}f_{k-i-1}^{(2)}}{(i+1)g(k-\ell-i-1)}\\
  &\qquad = c^1_{\ell k}\frac{q_{\ell+1,k}}{q_{\ell k}}
    \cdot\frac{g(k-\ell-1)}{\eta\,f_\ell^{(1)}\,g(k-\ell-i-1)}
  - c^2_{\ell k}\frac{q_{\ell,k+1}}{q_{\ell k}}\cdot\frac{e_{k+1}^{(2)}}{(i+1)\,g(k-\ell)}.
\end{split}
\end{equation}
The first condition holds since the recursion
relations~\eqref{eq_recursion} are satisfied. The second simplifies to
\begin{equation}
\begin{split}
  &a_{k-i}^{(2)}(u) 
    - d_{\ell+i+1}^{(1)}(u)\,\frac{\eta\,e_{k-i}^{(2)}f_{k-i-1}^{(2)}}{(i+1)g(k-\ell-i-1)}\\
  &\qquad = d_k^{(2)}(u)\,\frac{g(2\,\lambda_2-\ell)g(-\ell-1)}{g(k-\ell)g(k-\ell-i-1)}
  - d_\ell^{(1)}(u)\,\frac{\eta\,e_{k+1}^{(2)}f_k^{(2)}}{(i+1)\,g(k-\ell)}
\end{split}
\end{equation}
which can be verified in a direct computation.

The procedure to prove the assertion for $\Delta C(u)$ is similar. In
order to understand in detail how cocommutativity arises from the
choice of the coefficients $q_{\ell k}$ in~\eqref{eq_recursion}, it is
important to contrast the calculation for $\Delta B(u)$ with that for
$\Delta C(u)$. Thus we give the main steps in detail again.
Firstly,
\begin{equation}
\label{eq_dcnewbasis}
\begin{split}
  &\Delta C(u)\,v_{\ell k}\\
  &= q_{\ell k}\,\sum_{i=0}^kM^{k,\ell}_i\,\Delta C(u)\ket{\ell+i,k-i}\\
  &= q_{\ell k}\,\sum_{i=0}^kM^{k,\ell}_i\,\eta\,
        \bigl(a_{\ell+i}^{(1)}(u)e_{k-i}^{(2)}\ket{\ell+i,k-i-1}
         + e_{\ell+i}^{(1)}\,d_{k-i}^{(2)}(u)\ket{\ell+i-1,k-i}\bigr)\\
  &= q_{\ell k}\,\sum_{i=1}^kM^{k,\ell}_i\,\biggl(
         d_{k-i}^{(2)}(u)\,\eta e_{\ell+i}^{(1)}
         - a_{\ell+i-1}^{(1)}(u)\,\frac{i\,g(k-\ell-i)}{f_{\ell+i-1}^{(1)}}
         \biggr)\ket{\ell+i-1,k-i}\\
  &\qquad + q_{\ell k}\,\eta\,e_\ell^{(1)}d_k^{(2)}(u)\,\ket{\ell-1,k},
\end{split}
\end{equation}
again using the notation of~\eqref{eq_defineabbrev}. In the third line,
the summation index was shifted and Lemma~\ref{lemma_propertym} used. Writing
\begin{equation}
  c^3_{\ell k} = \eta\,d_k^{(2)}(u)\,e_\ell^{(1)}\,\frac{g(-\ell)}{g(k-\ell)},\qquad
  c^4_{\ell k} = \eta\,d_\ell^{(1)}(u)\,e_k^{(2)}\,\frac{g(k)}{g(k-\ell)},
\end{equation}
it remains to show that
\begin{equation}
\label{eq_dcnewbasis2}
\begin{split}
  &\Delta C(u)\,v_{\ell k}\\
  &= c^3_{\ell k}v_{\ell-1,k} + c^4_{\ell k}v_{\ell,k-1}\\
  &= c^3_{\ell k}q_{\ell-1,k}\,\sum_{i=0}^kM^{k,\ell-1}_i\ket{\ell+i-1,k-i}
    + c^4_{\ell k}q_{\ell,k-1}\,\sum_{i=0}^{k-1}M^{k-1,\ell}_i\ket{\ell+i,k-i-1}\\
  &= \sum_{i=1}^kM^{k,\ell}_i\,\biggl(
      c^3_{\ell k}q_{\ell-1,k}\frac{f_{\ell-1}^{(1)}\,g(k-\ell-i)}
                                  {f_{\ell+i-1}^{(1)}\,g(k-\ell)}\\
  &\qquad - c^4_{\ell k}q_{\ell,k-1}\frac{i\,g(k-\ell-1)}
                                         {\eta\,e_k^{(2)}f_{\ell+i-1}^{(1)}}
      \biggr)\ket{\ell+i-1,k-i} + c^3_{\ell k}q_{\ell-1,k}\ket{\ell-1,k}.
\end{split}
\end{equation}
In the last line, the summation index was shifted and, using
Lemma~\ref{lemma_propertym} several times, the coefficient
$M^{k\,\ell}_i$ was cast in the same form as
in~\eqref{eq_dcnewbasis}. Both expressions~\eqref{eq_dcnewbasis}
and~\eqref{eq_dcnewbasis2} are equal if and only if for all
$k,\ell\in\N_0$
\begin{equation}
  q_{\ell k}\,\eta e_\ell^{(1)}d_k^{(2)}(u) = c^3_{\ell k}q_{\ell-1,k}
\end{equation}
and for all $i\in\{0,\ldots,k\}$
\begin{equation}
\begin{split}
  &\eta\,d_{k-i}^{(2)}(u)\,e_{\ell+i}^{(1)}f_{\ell+i-1}^{(1)}
  - a_{\ell+i-1}^{(1)}(u)\,i\,g(k-\ell-i)\\
  &\qquad = c^3_{\ell k}\frac{q_{\ell-1,k}}{q_{\ell k}}
      \cdot\frac{f_{\ell-1}^{(1)}\,g(k-\ell-i)}{g(k-\ell)}
  - c^4_{\ell k}\frac{q_{\ell,k-1}}{q_{\ell k}}\cdot\frac{i\,g(k-\ell-1)}{\eta e_k^{(2)}}.
\end{split}
\end{equation}
The first condition holds since the recursion
relations~\eqref{eq_recursion} are satisfied. The second simplifies to
\begin{equation}
\begin{split}
  &\eta\,d_{k-i}^{(2)}(u)\,e_{\ell+i}^{(1)}f_{\ell+i-1}^{(1)}
  - a_{\ell+i-1}^{(1)}(u)\,i\,g(k-\ell-i)\\
  &\qquad = d_k^{(2)}(u)\,\frac{\eta e_\ell^{(1)}f_{\ell-1}^{(1)}\,g(k-\ell-i)}{g(k-\ell)}
  - d_\ell^{(1)}(u)\,\frac{i\,g(k)\,g(-2\,\lambda_1+k-1)}{g(k-\ell)},
\end{split}
\end{equation}
which can be verified in a direct computation.
\end{proof}

Writing the last proposition using operators, we obtain

\begin{corollary}
\label{cor_newcopro}
The twisted coproduct on $V_{\lambda_1}(\delta_1)\otimes
V_{\lambda_2}(\delta_2)$ can be expressed as follows:
\begin{mathletters}
\label{eq_new_copro}
\begin{eqnarray}
  F\cdot\Delta D(u)\cdot F^{-1} 
    &=& D(u)\otimes D(u),\\
  F\cdot\Delta B(u)\cdot F^{-1} 
    &=& B(u)\otimes D(u)\,\frac{\delta_1-\delta_2+\eta\,(\Ha+\lambda_2)}
                               {\delta_1-\delta_2+\eta\,(\Ha-\Hb)}\nn\\
      &+& D(u)\otimes B(u)\,\frac{\delta_1-\delta_2+\eta\,(-\lambda_1-\Hb)}
                               {\delta_1-\delta_2+\eta\,(\Ha-\Hb)},\\
  F\cdot\Delta C(u)\cdot F^{-1} 
    &=& C(u)\otimes D(u)\,\frac{\delta_1-\delta_2+\eta\,(\Ha-\lambda_2)}
                               {\delta_1-\delta_2+\eta\,(\Ha-\Hb)}\nn\\
      &+& D(u)\otimes C(u)\,\frac{\delta_1-\delta_2+\eta\,(\lambda_1-\Hb)}
                               {\delta_1-\delta_2+\eta\,(\Ha-\Hb)}.
\end{eqnarray}%
\end{mathletters}%
In particular the twisted coproduct is cocommutative.
\end{corollary}

\begin{remark}
\begin{myenumerate}
\item
  The results for the new coproduct~\eqref{eq_new_copro} are compared
  with the results of~\cite{Te99} in Section~\ref{sect_terras}.
\item
  Since the homomorphism of algebras $Y(\ssl_2)\to U(\ssl_2)$, which
  was used above to express the coproduct in terms of the Lie algebra
  generators $E$, $F$ and $H$, is not a homomorphism of co-algebras,
  the above expressions for the new coproduct do not allow an easy
  conclusion about the new coproduct on the Yangian.
\item
  Cocommutativity of $F\cdot\Delta A(u)\cdot F^{-1}$ follows
  immediately because the relation for the quantum
  determinant~\eqref{eq_qdet} can be solved for $A(u)$. Since $\qdet
  T(u)$ is group-like, the relation extends to the coproduct, and
  since it is central, we can apply $F\cdot(\cdot)\cdot F^{-1}$
  without changing the structure of the equation.
\end{myenumerate}
\end{remark}

Finally, the analogous construction can be made for the twist $\tilde
F_{12}$ which is obtained if $\Delta A(u)$ is diagonalized. We just
summarize the results. 

\begin{proposition}
If the coefficients $\tilde q_{\ell k}$ of $\tilde Q_{12}^{-1}$ satisfy the recursion
relations
\begin{equation}
\label{eq_recursiontilde}
  \frac{\tilde q_{\ell+1,k}}{\tilde q_{\ell k}} = 
    \frac{\tilde g(k-\ell)}{\tilde g(2\,\lambda_2-\ell)},\qquad
  \frac{\tilde q_{\ell k}}{\tilde q_{\ell,k-1}} = \frac{\tilde g(k)}{\tilde g(k-\ell)},
\end{equation}
where $\tilde g(x)=\delta_2-\delta_1+\eta\,(\lambda_1-\lambda_2+x)$,
then the action of the coproducts on the basis vectors $\tilde v_{\ell
k}$, see~\eqref{eq_eigenvecta} and~\eqref{eq_f12}, is given by
\begin{eqnarray}
  \Delta A(u)\,\tilde v_{\ell k} 
    &=& a_\ell^{(1)}(u)\cdot a_k^{(2)}(u)\,\tilde v_{\ell k},\\
  \Delta B(u)\,\tilde v_{\ell k} 
    &=& \eta f_\ell^{(1)}\,a_k^{(2)}(u)\,
             \frac{\tilde g(2\,\lambda_2-\ell)}
                  {\tilde g(k-\ell)}\tilde v_{\ell+1,k}
      + \eta f_k^{(2)}\,a_\ell^{(1)}(u)\,
             \frac{\tilde g(-2\,\lambda_1+k)}
                  {\tilde g(k-\ell)}\tilde v_{\ell,k+1},\\
  \Delta C(u)\,\tilde v_{\ell k} 
    &=& \eta e_\ell^{(1)}\,a_k^{(2)}(u)\,
             \frac{\tilde g(-\ell)}
                  {\tilde g(k-\ell)}\tilde v_{\ell-1,k}
      + \eta e_k^{(2)}\,a_\ell^{(1)}(u)\,
             \frac{g(k)}{\tilde g(k-\ell)}\tilde v_{\ell,k-1}.
\end{eqnarray}%
\end{proposition}

\begin{proof}
The proof is completely analogous to that of
Proposition~\ref{prop_newcoprod}. Here
Lemma~\ref{lemma_propertymtilde} is used to deal with the coefficients
$\tilde M^{k,\ell}_i$ and with the shifts of the summation index which are
necessary here.
\end{proof}

\begin{corollary}
The coproduct on $V_{\lambda_1}(\delta_1)\otimes
V_{\lambda_2}(\delta_2)$ twisted with the twist $\tilde F$
diagonalizing $\Delta A(u)$ can be expressed as follows:
\begin{mathletters}
\begin{eqnarray}
  \tilde F\cdot\Delta A(u)\cdot\tilde F^{-1} 
    &=& A(u)\otimes A(u),\\
  \tilde F\cdot\Delta B(u)\cdot\tilde F^{-1} 
    &=& B(u)\otimes A(u)\,\frac{\delta_2-\delta_1+\eta\,(\Ha+\lambda_2)}
                               {\delta_2-\delta_1+\eta\,(\Ha-\Hb)}\nn\\
      &+& A(u)\otimes B(u)\,\frac{\delta_2-\delta_1+\eta\,(-\lambda_1-\Hb)}
                               {\delta_2-\delta_1+\eta\,(\Ha-\Hb)},\\
  \tilde F\cdot\Delta C(u)\cdot\tilde F^{-1} 
    &=& C(u)\otimes A(u)\,\frac{\delta_2-\delta_1+\eta\,(\Ha-\lambda_2)}
                               {\delta_2-\delta_1+\eta\,(\Ha-\Hb)}\nn\\
      &+& A(u)\otimes C(u)\,\frac{\delta_2-\delta_1+\eta\,(\lambda_1-\Hb)}
                               {\delta_2-\delta_1+\eta\,(\Ha-\Hb)}.
\end{eqnarray}%
\end{mathletters}%
In particular it is cocommutative. Cocommutativity of $\Delta D(u)$
follows now using the quantum determinant.
\end{corollary}

\subsection{Towards a universal expression}

Finally, we would like to write down the operators $Q_{12}^{-1}$
resp.\ $\tilde Q_{12}^{-1}$ in a form that is independent of the
particular representation. Firstly the recursion
formulas~\eqref{eq_recursion} have to be solved. A case by case study
of these conditions for small finite-dimensional representations makes
it possible to find a solution:

\begin{lemma}
The coefficients 
\begin{equation}
\label{eq_recursionsolve}
  q_{\ell k} = \prod_{j=0}^{k-1}
    \frac{\delta_1-\delta_2+\eta\,(\lambda_1-\lambda_2-\ell+j)}
         {\delta_1-\delta_2+\eta\,(-\lambda_1-\lambda_2+j)}
\end{equation}
satisfy the recursion relations~\eqref{eq_recursion}.
\end{lemma}

Recall that
$g(x)=\delta_1-\delta_2+\eta\,(\lambda_1-\lambda_2+x)$. The proof is a
direct computation. For a universal expression, it is not desirable to
have a range of the product in~\eqref{eq_recursionsolve} which depends
on the index $k$. Here $k$ corresponds to the weight of the right
factor of the tensor product. The dependence on $k$ can be avoided if
one uses quotients of Gamma-functions. In the following we write
$\Ha\ket{\ell,k}=(\lambda_1-\ell)\ket{\ell,k}$
resp.~$\Hb\ket{\ell,k}=(\lambda_2-k)\ket{\ell,k}$:

\begin{proposition}
The expression
\begin{eqnarray}
\label{eq_q12i}
Q_{12}^{-1}=
  \frac{\Gamma\bigl((\delta_1-\delta_2)/\eta+\Ha-\Hb\bigr)}
       {\Gamma\bigl((\delta_1-\delta_2)/\eta+\Ha-\lambda_2\bigr)}\cdot
  \frac{\Gamma\bigl((\delta_1-\delta_2)/\eta-\lambda_1-\lambda_2\bigr)}
       {\Gamma\bigl((\delta_1-\delta_2)/\eta-\lambda_1-\Hb\bigr)},
\end{eqnarray}
specializes to the solution~\eqref{eq_recursionsolve} of the recursion
relations on all weight vectors of the representations
$V_{\lambda_1}(\delta_1)\otimes V_{\lambda_2}(\delta_2)$.
\end{proposition}

A completely analogous construction is available for the recursion
relations~\eqref{eq_recursiontilde} for the case where $\Delta A(u)$
is diagonalized. We just summarize the results:

\begin{lemma}
The coefficients 
\begin{equation}
\label{eq_recursionsolvetilde}
  \tilde q_{\ell k} = \prod_{j=1}^{\ell}
    \frac{\delta_2-\delta_1+\eta\,(\lambda_1-\lambda_2+k-\ell+j)}
         {\delta_2-\delta_1+\eta\,(\lambda_1+\lambda_2-\ell+j)}
\end{equation}
satisfy the recursion relations~\eqref{eq_recursion}.
\end{lemma}

Recall that $\tilde
g(x)=\delta_2-\delta_1+\eta\,(\lambda_1-\lambda_2+x)$. The proof is
again a direct computation.

\begin{proposition}
The expression
\begin{eqnarray}
\tilde Q_{12}^{-1}=
  \frac{\Gamma\bigl((\delta_2-\delta_1)/\eta+\lambda_1-\Hb+1\bigr)}
       {\Gamma\bigl((\delta_2-\delta_1)/\eta+\Ha-\Hb+1\bigr)}\cdot
  \frac{\Gamma\bigl((\delta_2-\delta_1)/\eta+\Ha+\lambda_2+1\bigr)}
       {\Gamma\bigl((\delta_2-\delta_1)/\eta+\lambda_1+\lambda_2+1\bigr)},
\end{eqnarray}
specializes to the solution~\eqref{eq_recursionsolvetilde} of the
recursion relations on all weight vectors of representations
$V_{\lambda_1}(\delta_1)\otimes V_{\lambda_2}(\delta_2)$.
\end{proposition}

%
\section{The universal $R$-matrix of the Yangian $Y(\ssl_2)$}
%
\label{sect_universal}

In this section we show how the factorizing twists can be used to
calculate $R$-matrices for the representations
$V_{\lambda_1}(\delta_1)\otimes V_{\lambda_2}(\delta_2)$.

\subsection{The Gauss decomposition}

Khoroshkin and Tolstoy~\cite{KhTo96} calculate the universal
$R$-matrix of the quantum double $\sym{D}Y(\ssl_2)$ of the Yangian
$Y(\ssl_2)$ and, exploiting the similarities of representations of
$\sym{D}Y(\ssl_2)$ with those of $Y(\ssl_2)$, obtain an expression for
the $R$-matrix that holds on generic evaluation representations of
$Y(\ssl_2)$. This $R$-matrix is presented in its Gauss decomposition
\begin{equation}
  R = \alignidx{R_+\, R_0\, R_-},
\end{equation}
see Theorem~5.1 in~\cite{KhTo96}, where the triangular parts $R_+$ and
$R_-$ simplify according to~(6.6) and~(6.7) in~\cite{KhTo96}. In order
to compare these results with our notation, we replace
$a\mapsto\delta_1/\eta$, $b\mapsto\delta_2/\eta$ and take into
account that the Lie algebra $\ssl_2$ in~\cite{KhTo96} is written in
a Chevalley basis, but in this paper in a Cartan-Weyl basis, \ie\
$h\mapsto 2H$. The result from~\cite{KhTo96} thus reads in our
notation
\begin{eqnarray}
R_+ &=& \sum_{n=0}^\infty\frac{\eta^n}{n!}E^n\otimes F^n \biggl(\prod_{j=1}^n
        {(\delta_1-\delta_2+\eta(\Ha-\Hb)+\eta\,j)}^{-1}\biggr),\\
R_- &=& \sum_{n=0}^\infty\frac{\eta^n}{n!}\biggl(\prod_{j=1}^n
        {(\delta_1-\delta_2+\eta(\Ha-\Hb)+\eta\,j)}^{-1}\biggr)
        F^n\otimes E^n.
\end{eqnarray}

For the specialization of the diagonal part $R_0$ to evaluation
representations, we write~(6.13) from~\cite{KhTo96} in our notation:
\begin{equation}
\label{eq_r0_full}
\begin{split}
  R_0\,\ket{\ell,k} 
    &= \frac{\Gamma\bigl(\frac{(\delta_1-\delta_2)/\eta+\lambda_1-\lambda_2+k+1}{2}\bigr)}
             {\Gamma\bigl(\frac{(\delta_1-\delta_2)/\eta+\lambda_1-\lambda_2+1}{2}\bigr)}
   \cdot\frac{\Gamma\bigl(\frac{(\delta_1-\delta_2)/\eta+\lambda_1-\lambda_2+k+2}{2}\bigr)}
             {\Gamma\bigl(\frac{(\delta_1-\delta_2)/\eta-\lambda_1+\lambda_2+1}{2}\bigr)}\\
&\times\frac{\Gamma\bigl(\frac{(\delta_1-\delta_2)/\eta-\lambda_1-\lambda_2+k}{2}\bigr)}
             {\Gamma\bigl(\frac{(\delta_1-\delta_2)/\eta+\lambda_1+\lambda_2+2}{2}\bigr)}
   \cdot\frac{\Gamma\bigl(\frac{(\delta_1-\delta_2)/\eta-\lambda_1-\lambda_2+k+1}{2}\bigr)}
             {\Gamma\bigl(\frac{(\delta_1-\delta_2)/\eta-\lambda_1-\lambda_2}{2}\bigr)}\\
&\times\frac{\Gamma\bigl(\frac{(\delta_1-\delta_2)/\eta+\lambda_1-\lambda_2-\ell}{2}\bigr)}
             {\Gamma\bigl(\frac{(\delta_1-\delta_2)/\eta+\lambda_1-\lambda_2+k-\ell}{2}\bigr)}
   \cdot\frac{\Gamma\bigl(\frac{(\delta_1-\delta_2)/\eta+\lambda_1-\lambda_2-\ell+1}{2}\bigr)}
             {\Gamma\bigl(\frac{(\delta_1-\delta_2)/\eta+\lambda_1-\lambda_2+k-\ell+1}{2}\bigr)}\\
&\times\frac{\Gamma\bigl(\frac{(\delta_1-\delta_2)/\eta+\lambda_1+\lambda_2-\ell+1}{2}\bigr)}
             {\Gamma\bigl(\frac{(\delta_1-\delta_2)/\eta+\lambda_1-\lambda_2+k-\ell+1}{2}\bigr)}
   \cdot\frac{\Gamma\bigl(\frac{(\delta_1-\delta_2)/\eta+\lambda_1+\lambda_2-\ell+2}{2}\bigr)}
             {\Gamma\bigl(\frac{(\delta_1-\delta_2)/\eta+\lambda_1-\lambda_2+k-\ell+2}{2}\bigr)}\,\ket{\ell,k}.
\end{split}
\end{equation}
In order to compare this with the results of
Sections~\ref{sect_triangular} and~\ref{sect_diagonal}, we need to
know how the product $\alignidx{F_{21}^{-1}F_{12}}$ of our twists is normalized
compared to~\eqref{eq_r0_full}. Since our twist acts as the identity
on the highest weight vector, we have to divide $R_0$
in~\eqref{eq_r0_full} by the character $\chi$ of the $R$-matrix
(see~(6.14) and the corresponding comments in~\cite{KhTo96}),
\begin{equation}
  \chi = \frac{\Gamma\bigl(\frac{(\delta_1-\delta_2)/\eta+\lambda_1+\lambda_2+1}{2}\bigr)}
              {\Gamma\bigl(\frac{(\delta_1-\delta_2)/\eta+\lambda_1-\lambda_2+1}{2}\bigr)}
    \cdot\frac{\Gamma\bigl(\frac{(\delta_1-\delta_2)/\eta-\lambda_1-\lambda_2+1}{2}\bigr)}
              {\Gamma\bigl(\frac{(\delta_1-\delta_2)/\eta-\lambda_1+\lambda_2+1}{2}\bigr)}.
\end{equation}
The character $\chi$ depends on the representations under study via
the highest weights $\lambda_j$. It is determined by the non-linear
relations in the definition of the quasi-triangular
structure~\eqref{eq_nonlineara} and~\eqref{eq_nonlinearb}. Since we do
not know the form of the coproduct applied to $R$, we cannot use these
conditions, and thus cannot determine $\chi$ from our calculation.

The quotient $R_0/\chi$ can finally be simplified using
\begin{equation}
  \frac{\Gamma(\frac{\alpha+m}{2})\Gamma(\frac{\alpha+m+1}{2})}
       {\Gamma(\frac{\alpha}{2})\Gamma(\frac{\alpha+1}{2})}
 \cdot\frac{\Gamma(\frac{\beta}{2})\Gamma(\frac{\beta+1}{2})}
           {\Gamma(\frac{\beta+m}{2})\Gamma(\frac{\beta+m+1}{2})}
 = \frac{\Gamma(\alpha+m)}{\Gamma(\alpha)}\cdot\frac{\Gamma(\beta)}{\Gamma(\beta+m)},
\end{equation}
where $m\in\N_0$, yielding
\begin{equation}
\label{eq_r0_simple}
\begin{split}
  R_0/\chi &=
        \frac{\Gamma\bigl((\delta_1-\delta_2)/\eta+\lambda_1-\Hb+1\bigr)}
             {\Gamma\bigl((\delta_1-\delta_2)/\eta+\Ha-\Hb+1\bigr)}
   \cdot\frac{\Gamma\bigl((\delta_1-\delta_2)/\eta+\Ha+\lambda_2+1\bigr)}
             {\Gamma\bigl((\delta_1-\delta_2)/\eta+\lambda_1+\lambda_2+1\bigr)}\\
&\times
        \frac{\Gamma\bigl((\delta_1-\delta_2)/\eta+\Ha-\lambda_2\bigr)}
             {\Gamma\bigl((\delta_1-\delta_2)/\eta+\Ha-\Hb\bigr)}
   \cdot\frac{\Gamma\bigl((\delta_1-\delta_2)/\eta-\lambda_1-\Hb\bigr)}
             {\Gamma\bigl((\delta_1-\delta_2)/\eta-\lambda_1-\lambda_2\bigr)}.
\end{split}
\end{equation}

\subsection{The factorization}

It is apparent that the diagonal part~\eqref{eq_r0_simple} is
factorized by the diagonal part of the twist:
\begin{equation}
\label{eq_factorr0}
  R_0/\chi = \alignidx{Q_{21}^{-1}Q_{12}}.
\end{equation}
The expression for $Q_{12}$ can be read off~\eqref{eq_q12i}. The
coefficients of $Q_{21}^{-1}$ are obtained
from~\eqref{eq_recursionsolve} exchanging $k\leftrightarrow\ell$,
$\delta_1\leftrightarrow\delta_2$ and
$\lambda_1\leftrightarrow\lambda_2$: 
\begin{equation}
\label{eq_q21ia}
  \prod_{j=0}^{\ell-1}
    \frac{\delta_2-\delta_1+\eta\,(\lambda_2-\lambda_1-k+j)}
         {\delta_2-\delta_1+\eta\,(-\lambda_2-\lambda_1+j)}
  =\prod_{j=0}^{\ell-1}
    \frac{\delta_1-\delta_2+\eta\,(\lambda_1-\lambda_2+k-\ell+1+j)}
         {\delta_1-\delta_2+\eta\,(\lambda_1+\lambda_2-\ell+1+j)},
\end{equation}
from which $Q_{21}^{-1}$ can be calculated
\begin{equation}
\label{eq_q21ib}
  Q_{21}^{-1} = \frac{\Gamma\bigl((\delta_1-\delta_2)/\eta+\lambda_1-\Hb+1\bigr)}
                     {\Gamma\bigl((\delta_1-\delta_2)/\eta+\Ha-\Hb+1\bigr)}
           \cdot\frac{\Gamma\bigl((\delta_1-\delta_2)/\eta+\Ha+\lambda_2+1\bigr)}
                     {\Gamma\bigl((\delta_1-\delta_2)/\eta+\lambda_1+\lambda_2+1\bigr)},
\end{equation}
confirming~\eqref{eq_factorr0}.

Finally from~\eqref{eq_f12i} we find $\alignidx{F_{21}^{-1}=R_+Q_{21}^{-1}}$ and
from~\eqref{eq_f12} that $\alignidx{F_{12}=Q_{12}R_-}$. This completes the
factorization
\begin{equation}
\label{eq_factorization}
  \alignidx{F_{21}^{-1}\,F_{12}}=\alignidx{R_+\,Q_{21}^{-1}\,Q_{12}\,R_-}=R
  =\alignidx{R_+\,(R_0/\chi)\,R_-}.
\end{equation}

Likewise for the twist $\tilde F$ which was obtained diagonalizing
$\Delta A(u)$, we have
\begin{equation}
  \tilde F_{12}^{-1} = {\left.F_{21}^{-1}\right|}_{\delta_1\leftrightarrow\delta_2},
\end{equation}
from which an analogous result follows.

The factorization of the diagonal part of the $R$-matrix
in~\eqref{eq_factorr0} is remarkably simple. In particular, it does
not involve any of the complications which have initially been
conjectured~\cite{Te99}. There it was speculated that the universal
forms of the two factors $\alignidx{F_{12}^{-1}\text{ and }F_{21}}$
might not in general be related by the usual inversion and swapping of
tensor factors, but only related in finite-dimensional
representations. Actually the factorization is simple for all highest
weight representations.

\subsection{A discrete symmetry}
\label{sect_terras}

There exists an alternative choice for the diagonal part of the twist
in addition to that constructed in Section~\ref{sect_diagonal}. In
Proposition~\ref{prop_newcoprod}, it is possible to use coefficients
$\hat q_{\ell k}$ instead of the $q_{\ell k}$ which satisfy alternative
recursion relations
\begin{equation}
\label{eq_alternaterec}
  \frac{\hat q_{\ell+1,k}}{\hat q_{\ell k}} = \frac{g(2\,\lambda_2-\ell)}{g(k-\ell)},\qquad
  \frac{\hat q_{\ell k}}{\hat q_{\ell,k-1}} = \frac{g(k-\ell)}{g(k)}.
\end{equation}
This leads to another twist $\hat F$ having the same triangular,
but a different diagonal part. The new coproduct differs from that in
Corollary~\ref{cor_newcopro} by the ordering of operators:
\begin{mathletters}
\label{eq_alternatecop}
\begin{eqnarray}
  \hat F\cdot\Delta B(u)\cdot\hat F^{-1} 
    &=& \frac{\delta_1-\delta_2+\eta\,(\Ha-\lambda_2)}
             {\delta_1-\delta_2+\eta\,(\Ha-\Hb)}\,B(u)\otimes D(u)\nn\\
    &+& \frac{\delta_1-\delta_2+\eta\,(\lambda_1-\Hb)}
             {\delta_1-\delta_2+\eta\,(\Ha-\Hb)}\,D(u)\otimes B(u),\\
  \hat F\cdot\Delta C(u)\cdot\hat F^{-1} 
    &=& \frac{\delta_1-\delta_2+\eta\,(\Ha+\lambda_2)}
             {\delta_1-\delta_2+\eta\,(\Ha-\Hb)}\,C(u)\otimes D(u)\nn\\
    &+& \frac{\delta_1-\delta_2+\eta\,(-\lambda_1-\Hb)}
             {\delta_1-\delta_2+\eta\,(\Ha-\Hb)}\,D(u)\otimes C(u).
\end{eqnarray}%
\end{mathletters}%
The recursion relations~\eqref{eq_alternaterec} are solved by
\begin{equation}
  \hat q_{\ell k}=\prod_{j=0}^{\ell-1}
    \frac{\delta_1-\delta_2+\eta\,(\lambda_1+\lambda_2-\ell+1+j)}
         {\delta_1-\delta_2+\eta\,(\lambda_1-\lambda_2+k-\ell+1+j)}.
\end{equation}
A comparison with~\eqref{eq_q21ia} and~\eqref{eq_q21ib} shows that the
corresponding diagonal operator
\begin{equation}
  \hat Q_{12}^{-1}\ket{\ell,k}=\hat q_{\ell k}\ket{\ell,k}
\end{equation}
is just $\alignidx{\hat Q_{12}^{-1}=Q_{21}}$. 

This shows that we are free to associate the factors
$\alignidx{Q_{21}^{-1}\text{ and }Q_{12}}$ of $R_0/\chi$
in~\eqref{eq_factorization} with either one of the triangular factors
to form a factorizing twist.  For $\hat F$ we have $\alignidx{\hat
F_{21}^{-1}=R_+Q_{12}}$ and $\alignidx{\hat F_{12}=Q_{21}^{-1}R_-}$
such that the factorization reads
\begin{equation}
  \alignidx{\hat F_{21}^{-1}\hat F_{12}} = \alignidx{R_+Q_{12}Q_{21}^{-1}R_-}
    = \alignidx{R_+\,(R_0/\chi)\,R_-}.
\end{equation}

The coproduct found in~\cite{Te99} using the modified Functional Bethe
Ansatz is the one given in~\eqref{eq_alternatecop}. The twist found in
that paper is our $\hat F$. We give in addition the explicit form of
its diagonal part $\alignidx{\hat Q_{12}^{-1}=Q_{21}}$ as the inverse
of~\eqref{eq_q21ib}.

\subsection{Existence of the factorizing twist in particular representations}
\label{sect_existence}

Finally, we comment on the question for which representations
$V_{\lambda_1}(\delta_1)\otimes V_{\lambda_2}(\delta_2)$ the twist
$F_{12}$ and its inverse exist. We restrict ourselves to the
finite-dimensional case where corresponding results for the existence
of the $R$-matrices are known. We recall that the twist is to
factorize the $R$-matrix in the relation
\begin{equation*}
  \Dop X(u) = R(\delta_1-\delta_2)\cdot\Delta X(u)\cdot
  R^{-1}(\delta_1-\delta_2).
\end{equation*}

\begin{remark}
\begin{myenumerate}
\item
  The $R$-matrix $R(\delta_1-\delta_2)$ as a function of
  $\delta_1-\delta_2$ is well-defined if and only if the representation
  $V_{\lambda_1}(\delta_1)\otimes V_{\lambda_2}(\delta_2)$ is
  irreducible. See Theorem~\ref{thm_irreducible}.
\item
  It is thus reasonable to expect that the twist exists at most in the
  irreducible representations, but there might be additional poles in
  the twist which cancel only in the product $\alignidx{F_{21}^{-1}\cdot
  F_{12}=R_{12}}$.
\end{myenumerate}
\end{remark}
In the following we analyze the structure of poles in
$\alignidx{F_{12}\text{ and }F_{12}^{-1}}$. We find that certain
factors from the denominator of the triangular part cancel with the
numerator of the diagonal part.

\begin{lemma}
The coefficients of $Q_{12}^{-1}$, see~\eqref{eq_recursionsolve}, can
be written
\begin{eqnarray}
\label{eq_qcancelled2}
  q_{\ell k} &=& \frac{\displaystyle
    \prod_{j=\max\{2\,\lambda_1-\ell+1,k+1\}}^{k+2\,\lambda_1-\ell}
      (\delta_1-\delta_2+\eta\,(-\lambda_1-\lambda_2+j-1))}
   {\displaystyle\prod_{j=1}^{\min\{k,2\,\lambda_1-\ell\}}
      (\delta_1-\delta_2+\eta\,(-\lambda_1-\lambda_2+j-1))}\\
\label{eq_qcancelled}
  &=& \prod_{j=1}^{\min\{k,2\,\lambda_1-\ell\}}
      \frac{\delta_1-\delta_2+\eta\,(\lambda_1-\lambda_2+k-\ell-j)}
           {\delta_1-\delta_2+\eta\,(-\lambda_1-\lambda_2+j-1)},
\end{eqnarray}
where numerator and denominator have no common factor.
\end{lemma}

\begin{proposition}
The expression $F_{12}^{-1}$, see~\eqref{eq_f12i}, is well-defined for
all finite-dimensional irreducible representations
$V_{\lambda_1}(\delta_1)\otimes V_{\lambda_2}(\delta_2)$.
\end{proposition}

\begin{proof}
First we show that all factors in the denominator of the triangular
part of $F_{12}^{-1}$ cancel with the numerator of
$Q_{12}^{-1}$. Poles can be present only if 
$F^n\otimes E^n\ket{\ell,k}\neq 0$, but this implies
$n\leq\min\{k,2\,\lambda_1-\ell\}$. The denominator of the triangular
part, see~\eqref{eq_f12i}, applied to $\ket{\ell,k}$ is
\begin{equation}
  \prod_{j=1}^n(\delta_1-\delta_2+\eta\,(\lambda_1-\lambda_2+k-\ell-j)).
\end{equation}
Since $n\leq\min\{k,2\,\lambda_1-\ell\}$, it cancels with the numerator
of $Q_{12}^{-1}$ as given in~\eqref{eq_qcancelled}.

The denominator of $Q_{12}^{-1}$ has a pole if and only if
\begin{equation}
  \frac{\delta_1-\delta_2}{\eta} = \lambda_1+\lambda_2-j+1,
\end{equation}
where $1\leq j\leq\min\{k,2\,\lambda_1-\ell\}$. But since $k\leq
2\,\lambda_2$ and $\ell\geq 0$, this implies $1\leq
j\leq\min\{2\,\lambda_1,2\,\lambda_2\}$ for which
$V_{\lambda_1}(\delta_1)\otimes V_{\lambda_2}(\delta_2)$ is reducible
(Theorem~\ref{thm_irreducible}).
\end{proof}

Similarly we can study $F_{12}$, see~\eqref{eq_f12}. In this case not
all factors in the denominator of the triangular part are cancelled.

\begin{lemma}
The expression $F_{12}$ in~\eqref{eq_f12} has poles at
\begin{equation}
  \frac{\delta_1-\delta_2}{\eta}=\lambda_1+\lambda_2-j+1,
\end{equation}
for precisely the values $j\in\{2,3,\ldots,2\,\lambda_1+2\,\lambda_2\}$.
\end{lemma}

\begin{proof}
The denominator of the triangular part of $F_{12}$ acting on
$\ket{\ell,k}$ is given by
\begin{equation}
\begin{split}
  &\prod_{j=1}^n(\delta_1-\delta_2+\eta\,(\lambda_1-\lambda_2+k-\ell-2n+j))\\
  &\qquad =\prod_{j=2+(2\,\lambda_1-\ell)+k-2n}^{1+(2\,\lambda_1-\ell)+k-n}
    (\delta_1-\delta_2+\eta\,(-\lambda_1-\lambda_2+j-1)),
\end{split}
\end{equation}
where $1\leq n\leq\min\{k,2\,\lambda_1-\ell\}$. Some factors cancel
with the numerator of $Q_{12}$ (see the expression for $Q_{12}^{-1}$
in~\eqref{eq_qcancelled2}). The remaining denominator of the
triangular part is
\begin{equation}
\label{eq_firstpoles}
  \prod_{j=\max\bigl\{2+(2\,\lambda_1-\ell)+k-2n,\min\{k+1,2\,\lambda_1-\ell+1\}\bigr\}}^{1+(2\,\lambda_1-\ell)+k-n}
    (\delta_1-\delta_2+\eta\,(-\lambda_1-\lambda_2+j-1)),
\end{equation}
since $2+(2\,\lambda_1-\ell)+k-2n>1$ and 
$1+(2\,\lambda_1-\ell)+k-n\geq\max\{k+1,(2\,\lambda_1-\ell)+1\}>
 \min\{k,2\,\lambda_1-\ell\}$.

The denominator of the diagonal part $Q_{12}$ is
\begin{equation}
\label{eq_secondpoles}
  \prod_{j=\max\{2\,\lambda_1-\ell+1,k+1\}}^{k+2\,\lambda_1-\ell}
    (\delta_1-\delta_2+\eta\,(-\lambda_1-\lambda_2+j-1)).
\end{equation}
Poles in $F_{12}$ can arise from both~\eqref{eq_firstpoles}
and~\eqref{eq_secondpoles}. If $k=0$ or $\ell=2\,\lambda_1$, then $n=0$,
and both products~\eqref{eq_firstpoles} and~\eqref{eq_secondpoles} are
empty.

If $k\geq 1$ and $\ell\leq 2\,\lambda_1-1$, there is an $n=1$
contribution to the product~\eqref{eq_firstpoles} which extends in this
case from $j=2$ to $j=(2\,\lambda_1-\ell)+k$. The second
product~\eqref{eq_secondpoles} runs from
$j=\max\{2\,\lambda_1-\ell+1,k+1\}\geq 2$ to
$j=(2\,\lambda_1-\ell)+k$. For arbitrary
$n\leq\min\{k,2\,\lambda_1-\ell\}$, these bounds are not exceeded.
\end{proof}

\begin{corollary}
The expression $F_{12}$ in~\eqref{eq_f12} is well-defined on
$V_{\lambda_1}(\delta_1)\otimes V_{\lambda_2}(\delta_2)$ if and only
if
\begin{equation}
  \frac{\delta_1-\delta_2}{\eta}\notin\{-\lambda_1-\lambda_2+1,
        -\lambda_1-\lambda_2+2,\ldots,\lambda_1+\lambda_2-1\}. 
\end{equation}
In particular there exist irreducible representations
$V_{\lambda_1}(\delta_1)\otimes V_{\lambda_2}(\delta_2)$ for which
$F_{12}$ does not exist.
\end{corollary}

\begin{remark}
According to Theorem~\ref{thm_irreducible}, the finite-dimensional
representation $V_{\lambda_1}(\delta_1)\otimes
V_{\lambda_2}(\delta_2)$ is irreducible if and only if
\begin{equation}
  \frac{\delta_1-\delta_2}{\eta}\notin
    \{-\lambda_1-\lambda_2,\ldots,-|\lambda_1-\lambda_2|-1;
      |\lambda_1-\lambda_2|+1,\ldots,\lambda_1+\lambda_2\},
\end{equation}
where the set does not include values around zero
$-|\lambda_1-\lambda_2|,\ldots,|\lambda_1+\lambda_2|$. For these
values there exists $F_{12}^{-1}$, but not $F_{12}$.

The simplest example is the representation $V_{\lambda}(\delta)\otimes
V_{\lambda}(\delta)$, where both $R(0)=P$ and $R^{-1}(0)=P$
are given by the operator which exchanges the tensor factors,
$P(a\otimes b)=b\otimes a$. Obviously,
\begin{equation}
  \Dop X(u) = P\cdot\Delta X(u)\cdot P.
\end{equation}
But $P$ is `too symmetric' to be factorized in an expression like
$P=\alignidx{F_{21}^{-1}\cdot F_{12}}$.
\end{remark}

\subsection{On tensor products of evaluation representations}

Having obtained the $R$-matrix on generic evaluation representations
$V_{\lambda_1}(\delta_1)\otimes V_{\lambda_2}(\delta_2)$, the
$R$-matrices on tensor products with more than two factors are
determined by the axioms
\begin{eqnarray}
  (\Delta\otimes\id)(R) &=& R_{13}R_{23},\nn\\
  (\id\otimes\Delta)(R) &=& R_{13}R_{12},\nn
\end{eqnarray}
of the quasi-triangular structure. It is thus possible to calculate
the $R$-matrices for tensor products of evaluation representations
even without knowing the action of the coproduct.

The $R$-matrices obtained this way, however, agree only up to the
character $\chi$ with the $R$-matrix which is obtained by representing
the universal $R$-matrix directly on a tensor product of evaluation
representations. The analysis in~\cite{KhTo96} shows however that the
characters $\chi$ are well understood so that this is not a serious
drawback.

Our method thus determines $R$-matrices for all tensor products of
evaluation representations. These include in particular all
finite-dimensional irreducible representations of $Y(\ssl_2)$ and
those representations of interest in applications to integrable
systems.

%
\section{Conclusion and outlook}
%
\label{sect_conclusion}

In this paper we have presented an elementary direct calculation of
the factorizing twist of $Y(\ssl_2)$ which is universal for all
evaluation representations. Having calculated the twist, it is
possible to recover the universal $R$-matrix specialized to a generic
evaluation representation. It appears automatically in a
canonical form, being Gauss decomposed as an upper triangular times a
diagonal times a lower triangular part.

The fact that this approach is successful underlines the importance of
studying the quantum groups like the Yangian $Y(\g)$ or the quantized
envelopes of the affine Lie algebras $U_q(\hat\g)$ in view of their
pseudo-triangularity. The factorizing twist seems more fundamental and
even more accessible than the universal $R$-matrix. This is very
relevant for the study of analogous constructions for Lie algebras
$\g$ of higher rank as well as for their applications to quantum
integrable systems. A thorough understanding of factorizing twists for
$\g$ of higher rank can be expected to simplify the nested version of
the Algebraic Bethe Ansatz dramatically as indicated
in~\cite{AlBo00}. 

At present the form of the factorizing twist on the abstract Yangian
algebra is not known. It may well be more difficult to deal with
than is the universal $R$-matrix because the twists exists for
fewer representations than $R$-matrices do. We thus expect only
a pseudo-twist.

However, in any case the factorizing twists provide the Yangian
$Y(\ssl_2)$ with an additional, very restrictive structure which has
not been fully exploited in the analysis of the algebra yet.

\acknowledgements

The author would like to thank DAAD for a scholarship
``Doktorandenstipendium im Rahmen des gemeinsamen
Hochschulsonderprogramms~III von Bund und L\"andern''. Thanks are also
due to A.~J.~Macfarlane, R.~Oeckl, V.~Terras, V.~B.~Kuznetsov and F.~Wagner for valuable
discussions and comments.  


\end{document}